\documentstyle[graphicx,multicol,prl,aps]{revtex}

\begin{document}

\draft

\title{On the Use of Optimized Monte Carlo Methods for
Studying Spin Glasses}

\author{E. Marinari$^1$, G. Parisi$^1$, F. Ricci-Tersenghi$^2$ and
F. Zuliani$^1$}

\address{
$^1$ Dipartimento di Fisica, INFN and INFM, 
Universit\`a di Roma {\em La Sapienza},\\
P. A. Moro 2, 00185 Roma, Italy.}

\address{
$^2$ The Abdus Salam International Center for Theoretical Physics,
Condensed Matter Group,\\
Strada Costiera 11, P.O. Box 586, I-34100 Trieste, Italy.}

\date{October $31$, $2000$}

\maketitle

\begin{abstract}  
We start from recently published numerical data by Hatano and
Gubernatis~\cite{HATGUB} to discuss properties of convergence to
equilibrium of optimized Monte Carlo methods (bivariate multi
canonical and parallel tempering).  We show that these data are not
thermalized, and they lead to an erroneous physical picture.  We shed
some light on why the bivariate multi canonical Monte Carlo method can
fail.
\end{abstract}


One of the main problems of numerical results originated from large
scale numerical simulations is that checking them is a task that is
frequently of the order of magnitude of checking a real experiment:
only repeating the full simulation, that demands availability of
computer time and codes, allows a full check of the results.

Here we will use as a starting point the work of reference
\cite{HATGUB} to discuss a few points both about optimized Monte Carlo
algorithms and about the behavior of $3D$ Edwards-Anderson (EA) spin
glasses in the low $T$ phase.  We will start by showing that the
numerical results reported in reference \cite{HATGUB}, as far as the
low $T$ values are concerned, are wrong: they are not equilibrium
averages over the Boltzmann probability. Because of that the physical
conclusions reached in the paper, supporting a trivial behavior of the
broken phase of $3D$ spin glasses, are wrong. On the contrary recent
numerical simulations \cite{RECENT} support, in this respect, a
behavior of the system consistent with the Replica Symmetry Breaking
(RSB) picture \cite{PARISI}. We will also shed some light on why the
optimized Monte Carlo method used in \cite{HATGUB} can fail.

In the following we will first analyze our numerical data obtained by
the {\em Parallel Tempering} Monte Carlo method \cite{PT}, focusing on
the analysis needed to establish that thermal equilibrium has been
reached \cite{OPTIMIZED}: we will use a large number of severe
criteria that ensure that thermalization has been reached.  After
showing that the results of \cite{HATGUB} are not correct in the low
$T$ region we will discuss some preliminary simulations done using the
same method used in \cite{HATGUB}, a bivariate version of the
Multi-Canonical Monte Carlo \cite{BERG_NEU}, and we will point out a
series of reasons for which a non careful implementation of this
strategy can fail.

\begin{figure}
\centering\includegraphics[width=0.6\textwidth,angle=0]{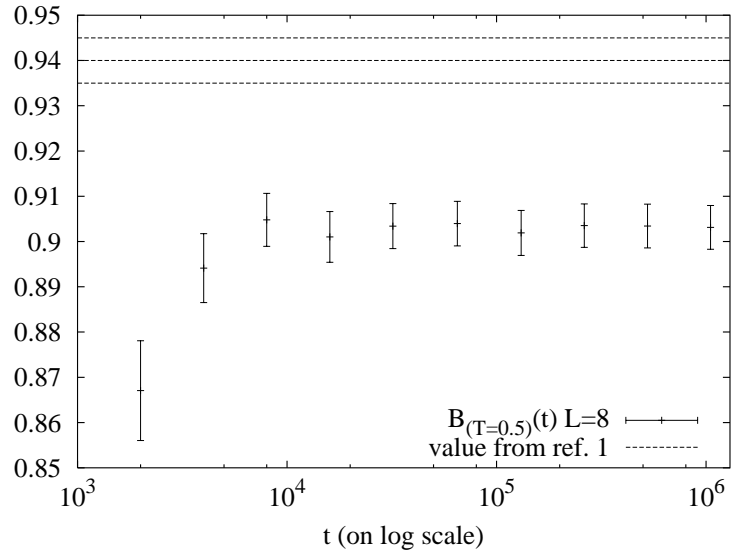}
\caption[a]{The Binder parameter, $B(t)$, averaged over logarithmic
time windows, as a function of time, at $T=0.5$.}
\protect\label{F-05}
\end{figure}

Let us start from our numerical data obtained through parallel
tempering\footnote{For sake of a complete reliability and without fear
of appearing over cautious we have chosen to rewrite all our codes in
a double blind pattern, with two different sets of programmers, using
different programming languages and different random number
generators: they always give statistically compatible results.}.  We
have simulated a $3D$ Edwards-Anderson spin glass, with binary random
quenched couplings, linear size $L=8$ (the largest size used in
\cite{HATGUB}), down to $T=0.5\simeq 0.5\, T_c$: let us note that in
our simulations for the same $T$ values we are able to thermalize
reliably lattices up to $L=16$, and that we just discuss here results
about the $L=8$ lattice, where we are completely confident about
thermalization, only because this is the largest lattice studied in
\cite{HATGUB}. We use a minimum value of the temperature
$T_{\mbox{min}}=0.5$, a number of temperatures $N_T=49$ and a constant
temperature step $\delta T = \frac{1}{30}$.  The measured correlation
times are always smooth functions of $T$ and no anomalies are
detected.

Our data at high $T$ turn out to be statistically compatible with the
ones of \cite{HATGUB}: in the high $T$ region there are no problems.

In figure \ref{F-05} we plot the value of the Binder parameter,

\begin{equation}
  B(t)\equiv\frac12\left(3-
  \frac
  {  \overline{\langle q^4(t)\rangle}    }
  { {\overline{\langle q^2(t)\rangle}}^2 }
  \right)\ ,
\end{equation}
averaged over logarithmic time windows, as a function of time at
$T=0.5$ (close to $0.5\,T_c$).  Averaging over logarithmic windows is
the safe approach to check convergence in time. We first average over
the last half of the total time extent of the run: this is the last
point on the right of the plot. We subdivide in the same way the other
half of the data, and the second point on the right is the average
over the second half of this time span: we continue in this way till
the origin of our Monte Carlo run. With a straight line we plot the
asymptotic data from \cite{HATGUB} as extracted from figure $7$ in the
paper (since we were estimating by hand we have been generous on the
statistical error): here there is no time dependence, we only plot
with a straight line the asymptotic value. The discrepancy of our data
and the data of \cite{HATGUB} is very large and statistically very
significant: definitely not an accident.

\begin{figure}
\centering\includegraphics[width=0.6\textwidth,angle=0]{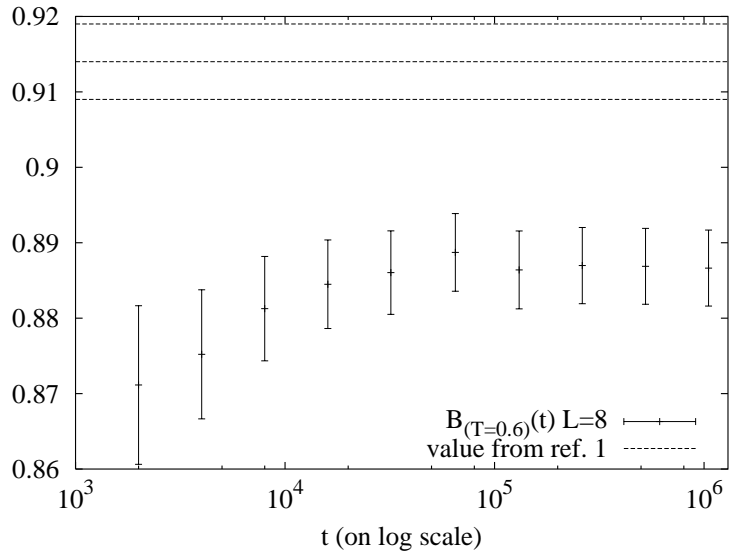}
\caption[a]{As in figure \ref{F-05}, but for $T=0.6$.}
\protect\label{F-06}
\end{figure}

In figure \ref{F-06} we plot the $T=0.6$ data from the same run,
always for the Binder parameter averaged over logarithmic time
windows: here $T$ is higher, and one could feel safer about
thermalization, but again there is a clear and significant discrepancy
among our data and the ones of \cite{HATGUB}. The dramatic stability
of our data for $B(t)$ at low $T$ is already a very good indicator of
a high level of thermalization. The results are stable at least during
the last eight subdivisions of our two million step runs, i.e. at least
from times going from $10^4$ to $2\cdot10^6$.

In order to be sure we are not trapped in some metastable situation we
have to check standard criteria about convergence, that in the case of
optimized dynamics can be quite difficult to assert
\cite{OPTIMIZED}. Let us note for example that in recent numerical
simulations \cite{RECENT} a careful discussion shows that weaker
criteria can be sufficient to guarantee thermalization, making in this
way possible to simulate more disorder sample with the same amount of
computer time (since one needs less thermal sweeps per sample). Here,
since thermalization is the main issue, we will check all of the most
stringent criteria.

First of all we have checked the acceptance rates of the tempering
sweeps in temperature: a bad choice of the $T$ values can make the
swap of the temperature value too rare. In our case the rates are very
high, of the order of $.7$ in all the temperature range: our parallel
tempering scheme is performing very well.

Secondly we have checked, as customary, if all configurations
(we have, as we said, $49$ of them) have spent a similar amount of
time in each one of the $49$ allowed $T$ values. This criterion is
important, since the first one could not be sufficient: spin
configurations could be spending time swapping 
among neighboring
$T$ values locally,
but never leave the high or the low $T$ region. Our {\em permanence
histograms} are very good: because of the large time extent of the
runs all configurations have visited all regions of the $T$ phase
space, and the permanence histograms are very flat. Again, this
is a powerful test of thermalization.

The last point we have checked is the symmetry under the exchange
$q\to-q$ of the $P_J(q)$ for the {\em individual} samples. Since the
overall flip of all spins is supposed to be a very slow mode of the
dynamics, once we have good statistics on this mode we expect to have
reached all the relevant regions of the phase space. Again, the
symmetry is excellent for all individual samples (even for the more
complex samples where the $P_J(q)$ has a non-trivial structure).

We consider this body of evidence as clear: our data are thermalized,
the numerical data hint evidence in favor of the RSB picture (as
confirmed by the data of \cite{RECENT}, where even at very low $T$
values one sees that $P(0)$ does not depend on $L$) and the method
used in \cite{HATGUB} did not allow a proper thermalization.

In order to get a better understanding of the situation, and some
hints about the reason of the failure of \cite{HATGUB} we have
implemented a code for rerunning their bivariate multi-canonical
simulations.

Our simulations closely follows the description given in the Appendix
of reference \cite{HATGUB} and by Hatano himself \cite{HATANO}.  The
analysis of few samples of sizes $L=4$, $6$, $8$ has been sufficient
in order to understand where the thermalization problems may come
from.  Unless differently specified we have always used $10^6$ Monte
Carlo Sweeps (MCS) for thermalizing and $10^7$ MCS for taking
measurements in {\em each} multi-canonical cycle.  The same number of
MCS has been used by the authors of \cite{HATGUB} only for $L=10$
\cite{HATANO} (less iterations have been used for smaller lattice
sizes).

The most delicate point during the thermalization process is the role
played by the {\em entropic barriers} during the multi-canonical
simulation.  In a model which undergoes a first order transition the
slowing down of the simulation at the critical point is essentially
due to the presence of a huge {\em energetic barrier} between the two
free energy minima. In this case the multi-canonical simulation works
fine \cite{BERG_NEU}, and it rapidly converges towards a regime where
every energy is sampled with the right probability, i.e. uniformly.
Problems may arise when the multi-canonical method is applied to spin
glasses or in general to models where entropic barriers play a central
role. To this respect the study of its performances in models with
only entropic barriers (e.g. backgammon model \cite{BACKGAMMON}) would
be illuminating.

\begin{figure}
\centering\includegraphics[width=0.6\textwidth,angle=0]{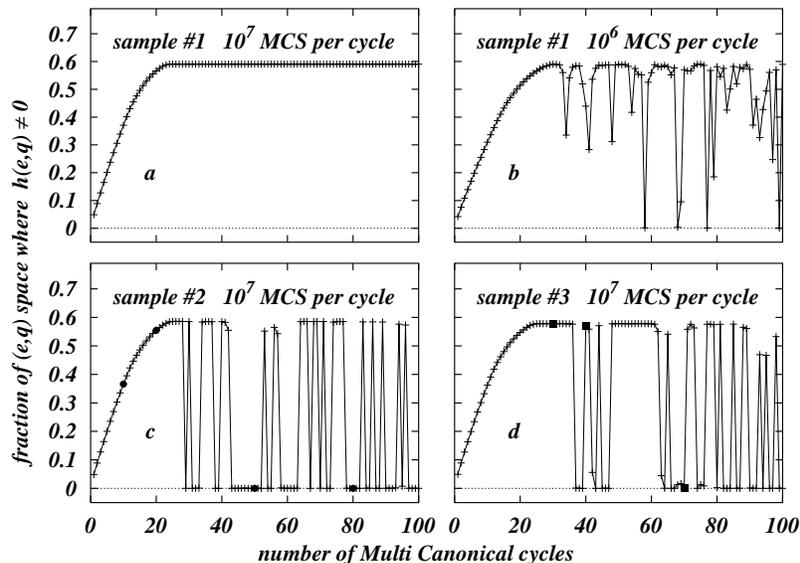}
\caption[a]{The fraction of $(e,q)$ space where the histogram $h(e,q)$
is different from zero as a function of the multi-canonical cycle
number.  Even for a very small system ($L=6$) strong convergence
problems arise.}
\label{F-FRAZ}
\end{figure}

\begin{figure}
\centering\includegraphics[width=0.6\textwidth,angle=0]{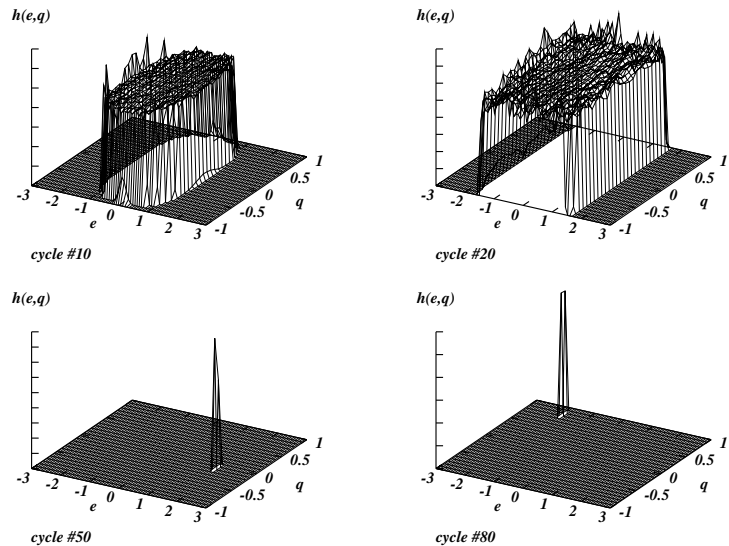}
\caption[a]{The evolution of the histogram $h(e,q)$ as a function of
multi-canonical cycles (sample \#2 in figure \ref{F-FRAZ}).}
\label{F-ISTO}
\end{figure}

Let us focus now specifically on the $3D$ EA model, and see how the
estimated density of states (DoS), $D(e,q)$, converges to the exact
one.  In particular we are interested in the histogram $h(e,q)$ which
counts the number of times, during a multi-canonical cycle, the system
is in a macroscopic state $(e,q)$ with energy $e$ and overlap $q$.
Thermalization is achieved when $h(e,q)$ is flat and much larger than
$1$ for all the physically allowed pairs $(e,q)$.  Starting from a
flat DoS, the region where $h(e,q) \gg 1$ broadens with the number of
multi-canonical cycles and eventually reaches the boundaries of the
allowed domain, $e \in [-e_0,e_0] \; q \in [-1,1]$, where $-e_0$ is
the ground-states energy (see the first two snapshots in figure
\ref{F-ISTO}, that we will discuss in better detail later on).  In
order to describe quantitatively the histogram evolution we plot in
figure \ref{F-FRAZ} the fraction of the $(e,q)$ space where $h(e,q)
\neq 0$, that is the fraction of macroscopic $(e,q)$ configurations
visited by the system during a multi-canonical cycle.  We expect this
fraction to increase more or less linearly during the first
multi-canonical cycles and then to reach a plateau when simulation is
thermalized (see figure \ref{F-FRAZ}.a, where things look good).  For
all the $L=4$ samples simulated we have observed this correct
behavior.  On the contrary for the $L=6$ samples, problems arise.  At
first, if the number of MCS is not large enough the simulation does
not converge at all.  In figure \ref{F-FRAZ}.b we show the results for
the same sample shown in figure \ref{F-FRAZ}.a, with the only
difference that $10^6$ MCS were used instead of $10^7$: here
thermalization problems are evident, since in some situations the
system simply gets trapped in a very small region of the phase space.
In different samples we have found analogous problems also when using
$10^7$ MCS (see figures \ref{F-FRAZ}.c and \ref{F-FRAZ}.d).  With
$10^6$ MCS the parallel tempering method is able to thermalize samples
up to $L=8$ for temperatures down to $T=0.3$ (for example at the
lowest $T$ value the Binder parameter thermalizes in $10^6$ MCS):
the bivariate multi-canonical method does not seem to be very
efficient for spin glasses.

In figure \ref{F-ISTO} we show the histogram evolution for sample \#2
(the same used in figure \ref{F-FRAZ}.c).  The four snapshots
correspond to the black dots in figure \ref{F-FRAZ}.c and clearly show
that the system, after reaching an apparently thermalized state with a
flat and broad $h(e,q)$, instead of keeping it for all subsequent
times, gets trapped in very small regions of the $(e,q)$ space (the
third and fourth snapshots in figure \ref{F-ISTO}).

How can we explain this behavior? During the first multi-canonical
cycles the dynamics of the system in the $(e,q)$ space is diffusive in
character, while when approaching the boundaries of the $e-q$ plane
(especially the energy ones) the system often gets trapped for very
long times.  The end of the diffusive behavior near to the ground
states can be easily explained in terms of accessibility, that is the
probability of decreasing the energy when the system is in a $(e,q)$
configuration and it makes a random move to a neighbor configuration.
For not too low energies the accessibility is high: in this case a
random walk in the configuration space corresponds to a random walk
in the $(e,q)$ space, which is a projection of the previous one.  On
the contrary for energies close to the one of the ground states the
accessibility is very low, due to the presence of a large number of
higher local minima. For example if the system is at the bottom of a
valley in the space of microscopic configurations, in order to further
decrease its energy (a little step in the macroscopic $(e,q)$ space)
it may need a long time, the time to find a deeper valley.  The
dynamics turns out to be strongly constrained for energies close to
the boundaries.

Having in mind that the dynamics becomes slower and slower close to
the energy boundaries, one can easily explain the peaks in figure
\ref{F-ISTO}.  The system firstly relaxes in a uniform way on a large
part of the $(e,q)$ space, the more accessible one.  Still many
allowed $(e,q)$ values are unvisited (because of the low
accessibility), their DoS estimation becomes very small and their
corresponding weights, $W(e,q) = 1 / D(e,q)$, huge.  When the system
reaches one of this configurations it can not leave it until the end
of the multi-canonical cycle, when $W(e,q)$ will be updated again.

In order to improve the convergence we have also tried to start with a
DoS estimated from the one of a thermalized $L=4$ sample.  The
convergence seems to be faster, however the problems giving rise to
the peak structure in the histogram remain unaltered.

Given that the thermalization task appears to be very hard, one should
at least try to use all thermalization checks available.  For example
the one based on the symmetry of the overlap distribution for every
sample, $P_J(q)$ should always be carefully checked: this analysis is
lacking in \cite{HATGUB}.

\begin{figure}
\centering\includegraphics[width=0.6\textwidth,angle=0]{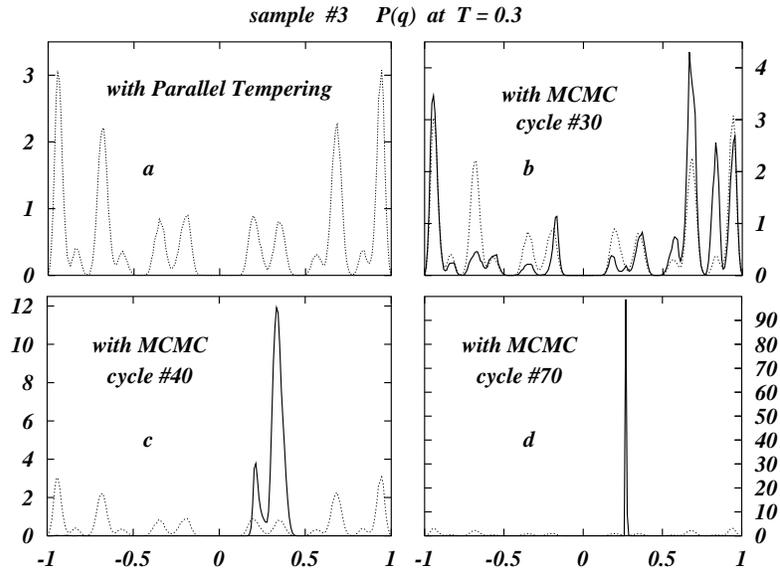}
\caption[a]{For a given $L=6$ sample (sample \#3 in figure \ref{F-FRAZ}) the
$P(q)$ measured with parallel tempering (top left) is symmetric, while
it may become much more narrow when a multi-canonical method is
employed.}
\label{F-PQ}
\end{figure}

In figure \ref{F-PQ} we show the overlap distribution $P_J(q)$ for the
single $L=6$ sample considered in figure \ref{F-FRAZ}.d at a low
temperature $T=0.3$ (these data come from a further parallel tempering
simulation, pushed to lower $T$ values).  In figure \ref{F-PQ}.a we
show the $P(q)$ measured with a parallel tempering simulation.  Its
very accurate symmetry is a strong evidence of complete
thermalization.  In the next $3$ plots (b,c and d) we show with
continuous lines the $P(q)$ measured with the multi-canonical method
(the chosen times correspond to the dots in figure \ref{F-FRAZ}.d).
We always superimpose the thermalized $P(q)$ for comparison.  It is
clear that, in the best case (see figure \ref{F-PQ}.b), the
multi-canonical method is not able to give results as good as the
parallel tempering does: in the worst cases it just gives a completely
wrong $P_J(q)$, with a single or a double peak.  The system may very
easily get stuck somewhere, and in these cases the estimated $P(q)$
would look much narrower than the correct one (see figure \ref{F-PQ}.c
and figure \ref{F-PQ}.d): measurements taken in such a biased
situation hint for a fake evidence in favor of a single peak $P(q)$,
and consequently of the droplet scenario.

\begin{figure}
\centering\includegraphics[width=0.6\textwidth,angle=0]{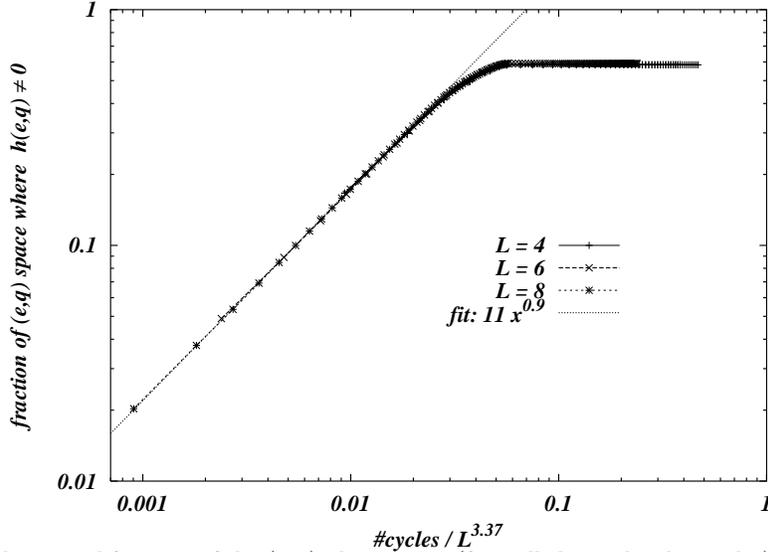}
\caption[a]{The scaling of the visited fraction of the $(e,q)$ phase
space (for well thermalized samples) shows that the equilibration time
must grow with the system size faster than $\tau \propto N^{1.1}$.}
\label{F-SCALING}
\end{figure}

As a last piece of evidence we consider the samples where the
bivariate multi canonical has been well behaved: the scaling of the
visited fraction of the $(e,q)$ phase space (for well thermalized
samples) reported in figure \ref{F-SCALING} supports the picture of a
diffusion-like evolution of the histogram.  The area of support of the
histogram grows more or less linearly with the number of
multi-canonical cycles (the best exponent estimate is 0.9).  Moreover,
the time for reaching the plateau (equilibration time) grows with
$\tau \propto L^{3.37} \propto N^{1.12}$, which seems to be very close
to the theoretical lower bound ($\tau \propto N$).  However this
result would hold {\em only if} the number of MCS per multi-canonical
cycle necessary for a proper thermalization is independent from the
system size $N$.  As we have already seen this is not true.  Indeed,
using the same $10^7$ MCS per multi-canonical cycle, the fraction of
well thermalized samples we have obtained is 100\% for $L=4$, around
40\% for $L=6$ and 0\% for $L=8$.  Because the requested number of MCS
per multi-canonical cycle grows with $N$ (apparently very fast), our
conclusion is that $\tau$ grows much faster than $N$ (simple arguments
by Berg \cite{BERG} suggest at least as $N^2$).

Concluding, we have seen how difficult it is to bring a bivariate
multi-canonical simulation of spin glasses to equilibrium and,
consequently, one possible reason of the failure of \cite{HATGUB} to
thermalize for $L=8$ (we have checked the failure of thermalization
with independent parallel tempering simulations).  When we say that
the simulation is not thermalized we mean that we can not use the
resulting DoS\footnote{Note that the DoS estimation actually used in
the measurements in \cite{HATGUB} is $D(e,q) h(e,q)$ and so it is
strongly affected by non-uniformities in the histogram.}  in order to
estimate the observables averages at all the temperature.  In
particular, as long as the simulation does not visit many times the
ground-states, we cannot believe to have enough information on the
ground-states structure.  However it may perfectly be that, after a
certain number of multi-canonical cycles, the estimated DoS gives good
averages at higher temperatures, which do not change if new low energy
states are reached.  We believe this is the case in \cite{HATGUB},
where data at not too low temperatures are perfectly compatible with
the ones obtained in previous work and fit the RSB scenario.

We thank N. Hatano for an useful correspondence regarding the
bivariate method.

\end{document}